\documentstyle[twocolumn,seceq]{jpsj}

\def\runtitle{Spin Dynamics of the 1D $J$-$J'$ Model and Spin-Peierls
Transition in CuGeO$_3$}
\def\runauthor{Hisatoshi {\sc Yokoyama} and Yasuhiro {\sc Saiga}}

\title
{
Spin Dynamics of the One-Dimensional $J$-$J'$ Model and
Spin-Peierls Transition in CuGeO$_3$
}

\author
{
Hisatoshi {\sc Yokoyama}\footnote{E-mail: yoko@cmpt01.phys.tohoku.ac.jp}
and Yasuhiro {\sc Saiga}
}

\inst
{
Department of Physics, Tohoku University,
Aramaki Aoba, Aoba-ku, Sendai 980-77\\
}

\recdate
{
\today
}

\def\etal{{\it et\ al.}\ }
\def\etalp{{\it et\ al}.\ }
\newcommand{\lsim}
 {\ \raise.35ex\hbox{$<$}\kern-0.75em\lower.5ex\hbox{$\sim$}\ }
\newcommand{\gsim}
 {\ \raise.35ex\hbox{$>$}\kern-0.75em\lower.5ex\hbox{$\sim$}\ }
\def\journal #1#2#3#4{#1 {\bf #2} (#4) #3}
\def\PR{Phys.\ Rev.}
\def\PRB{Phys.\ Rev.\ B}
\def\PRL{Phys.\ Rev.\ Lett.}

\def\BAPS{Bull.\ Am.\ Phys.\ Soc.}
\def\SSC{Solid State Commun.}

\def\PL{Phys.\ Lett.}
\def\PLA{Phys.\ Lett.\ A}

\def\JAP{J.\ Appl.\ Phys.}
\def\JMP{J.\ Math.\ Phys.}

\def\JPA{J.\ Phys.\ A}
\def\JPC{J.\ Phys.\ C}

\def\JPCS{J.\ Phys.\ Chem.\ Solids}

\def\JPSJ{J.\ Phys.\ Soc.\ Jpn.}
\def\MPLB{Mod.\ Phys.\ Lett.\ B}

\def\PTP{Prog.\ Theor.\ Phys.}

\abst
{
Spin dynamics as well as static properties of the one-dimensional
$J$-$J'$ model ($S=1/2$, $J>0$ and $0\le \alpha=J'/J\le 0.5$) are
studied by the exact diagonalization and the recursion method
of finite systems up to 26 sites.
Especially, the dynamical structure factor $S(q,\omega)$ is
investigated carefully for various values of $\alpha$.
As $\alpha$ increases beyond the gapless-gapful critical value
$\alpha_{\rm c}=0.2411$, there appear features definitely different
from the Heisenberg model but the same with the Majumdar-Ghosh model.
Some of these features depend only on the value of $\alpha$ and
not on $\delta$: a parameter introduced for the coupling alternation.
By comparing these results with a recent inelastic neutron scattering
spectrum of an inorganic spin-Peierls compound CuGeO$_3$
[M.\ Arai \etal: \journal{\PRL}{77}{3649}{1996}], it is found that
the frustration by $J'$ in CuGeO$_3$ is unexpectedly strong
($\alpha=0.4$-0.45), and at least $\alpha$ must be larger than
$\alpha_{\rm c}$ to some extent.
The value of $J$ is evaluated at $\sim 180$K consistent with
other estimations.
The coupling alternation is extremely small.
This large frustration is a primary origin of the various anomalous
properties CuGeO$_3$ possesses.
For comparison we refer also to $\alpha'$-NaV$_2$O$_5$.
}

\kword
{
spin dynamics, neutron scattering, spin-Peierls transition, CuGeO$_3$,
$J$-$J'$ model, Heisenberg model, Haldane-Shastry model,
Majumdar-Ghosh model, exact diagonalization, recursion method
}

\begin{document}
\sloppy
\maketitle

\section{Introduction}

In connection with the quasi spin gap found in the high-$T_{\rm c}$
superconductors, materials with a spin gap have been intensively studied
both by experiment and theory.
For a systematic understanding of the spin gap, it is indispensable to
investigate thoroughly the properties of basic models.
Among such models, the $S=1/2$ one-dimensional (1D) antiferromagnetic ($J>0$)
Heisenberg model with frustration by the next-nearest-neighbor exchange,
which we call the $J$-$J'$ model,\cite{MG}
\begin{equation}
{\cal H}=J\sum_j {\mib S}_j\cdot{\mib S}_{j+1}
        +J'\sum_j {\mib S}_j\cdot{\mib S}_{j+2},
\end{equation}
has some unique points.
It is well-known that this model changes its character
in the low-energy excitation from a gapless feature of the
Heisenberg model ($\alpha=J'/J=0$) to a finite-gap excitation
like the Majumdar-Ghosh (MG) case ($\alpha=0.5$) at
$\alpha_{\rm c}=0.2411$.\cite{ON}
At the MG point the degenerate ground states are exactly known as
products of the nearest-neighbor singlet wave functions.\cite{MG,SS}
So far, some static and thermodynamic properties have been elucidated
by numerical studies\cite{TH} and low-energy properties by the
conformal field theory.\cite{CFT,Chitra}
However, systematic studies of spin dynamics have been lacking.\cite{Yu}
\par

Another important aspect of this model is a close relationship with
the first inorganic spin-Peierls (SP) compound CuGeO$_3$.\cite{Hase}
The origin of the spin gap in CuGeO$_3$ is now established
as the lattice dimerization,\cite{Pouget,Hirota1} although it is
very small.
Some properties below the SP transition temperature $T_{\rm SP}\sim$14K
are common with the organic SP compounds and can be broadly understood
by the theory of Cross and Fisher.\cite{CF}
On the other hand, there are many unique aspects contradictory
to this theory.
For instance, phonon softening has not yet been observed,\cite{Kuroe}
which has been necessarily discovered for the organic SP compounds
as a driving force of the SP transition.
Moreover, the thermodynamic properties observed above $T_{\rm SP}$
do not coincide with the results of the simple Heisenberg model.\cite{Hase}
\par

Later, it was found that the latter contradiction can be reconciled
by introducing frustration due to the next-nearest-neighbor exchange
$J'$.\cite{Riera,Castilla}
Although many following theoretical studies\cite{followers,Poil1,Bouzerar}
have unanimously confirmed the importance of this frustration to explain
the experiments, the ratio of frustration $\alpha$ remains controversial
among them.
Originally, Riera and Dobry\cite{Riera} concluded $\alpha=0.36$ by fitting
the spin susceptibility $\chi(T)$.
Meanwhile, Castilla \etal\cite{Castilla} excluded the possibility for
$\alpha>\alpha_{\rm c}$ by a conjecture based on the theory of
Cross and Fisher and the dependence on temperature of the gap, and
concluded $\alpha=0.24$, which is a marginal value for the gapless
condition.
At any rate, the main point of the controversy is how the gap by $J'$
for $\alpha>\alpha_{\rm c}$ is compatible with the SP transition.
\par

In this paper, first we investigate spin dynamics of the
1D $J$-$J'$ model without coupling alternation carefully for various
values of $\alpha$ $(0\le\alpha\le 0.5)$ by the exact diagonalization
and the recursion method at zero temperature.\cite{GB}
As a result, the dynamical structure factor $S(q,\omega)$ has qualitatively
different characters, according as $\alpha<\alpha_{\rm c}$ or
$\alpha>\alpha_{\rm c}$; for $\alpha\sim\alpha_{\rm c}$ it resembles
that of the Haldane-Shastry (HS) model.
\par

Second, we compare these results with a complete spectrum of the
inelastic neutron scattering\cite{Arai} to search proper parameters
extensively in the $\alpha$-$\delta$ space, by adding the effect of
coupling alternation $\delta$ to the $J$-$J'$ model eq.\ (1.1).
Note that our fit is carried out at low temperatures
($T\lsim T_{\rm SP}$) in contrast with the previous ones by
$\chi(T)$,\cite{Riera,Castilla} which were done for $T>55K$.
Consequently, we have found that $\alpha$ thus obtained is unexpectedly
large ($\alpha=0.4$-0.45), and that at least $\alpha$ must be
somewhat larger than $\alpha_{\rm c}$.
This conclusion is drawn from the fact that the experimental spectrum
shows some evident features characteristic of the region
$\alpha>\alpha_{\rm c}$, which do not appear only by adding $\delta$
to eq.\ (1.1) with $\alpha<\alpha_{\rm c}$.
\par

Once $\alpha$ is specified, the value of $J$ is determined by
neutron data\cite{Nishi} as $\sim 180$K, which is consistent with
those obtained by other means.\cite{Riera,Nojiri}
On the other hand, $\delta$ is found to be very small (0.001-0.005).
We are convinced that the main origin of most unique properties
in CuGeO$_3$ is nothing but this large frustration, which is
not in competition but in concert with the lattice dimerization
in making a gap.
In this connection, we refer also to the second inorganic SP compound
$\alpha'$-NaV$_2$O$_5$, in which $\alpha$ seems small.
\par

This paper is organized as follows: In \S2 we mention the model
and the method.
Section 3 concentrates on the results for the $J$-$J'$ model without
coupling alternation ($\delta=0$).
In \S3.1 the Heisenberg case is taken up as a reference.
In \S3.2 the regime of $\alpha\sim\alpha_{\rm c}$ is
considered in relation to the HS model.
In \S3.3 characteristics of the gap regime are explained.
Section 4 is assigned to the comparison with the experiments
of CuGeO$_3$.
In \S4.1 we compare $S(q,\omega)$ with a complete spectrum
of the neutron scattering to determine the value of $\alpha$.
In \S4.2 the values of $J$ and $\delta$ are considered.
In \S4.3 based on the parameters thus determined, we discuss
a couple of other properties observed in CuGeO$_3$ and
$\alpha'$-NaV$_2$O$_5$.
We summarize in \S5.
In Appendix A we describe the single-mode approximation for the
$J$-$J'$ model.
In Appendix B we give a proof of a pure single mode at $q=\pi/2$
for the exactly soluble case: $2\alpha+\delta=1$.
\setcounter{equation}{0}
\section{Model and Method}
In this section we summarize our model and method briefly.
In this paper, we consider the 1D antiferromagnetic Heisenberg
Hamiltonian with the next-nearest-neighbor exchange and
the coupling alternation:\cite{Kuboki}
\begin{equation}
{\cal H}=J\sum_j\Bigl\{\left[1-(-1)^j\delta\right]{\mib S}_j\cdot{\mib S}_{j+1}
        +\alpha{\mib S}_j\cdot{\mib S}_{j+2}\Bigr\},
\end{equation}
where we assume $\alpha, \delta\ge 0$ and
${\mib S}_{j+N}={\mib S}_{j}$ with $N$ being the number of sites and even.
\par

For $\alpha=\delta=0$ eq.\ (2.1) is reduced to the Heisenberg model,
for which various ground-state and thermodynamic properties
are exactly known by the Bethe Ansatz; quite recently the structure
of spin excitation at $T=0$ has been clarified by using quantum group,
etc.\cite{BB}
The ground-state phase diagram of eq.\ (2.1) in the $\alpha$-$\delta$
plane was given by the conformal field theory;\cite{CFT,Chitra}
the regime of gapless excitation is limited on the segment
$0\le \alpha\le\alpha_{\rm c}$ and $\delta=0$.
At $\alpha_{\rm c}$ marginally irrelevant operators vanish
(conformal invariant), so that there exists no logarithmic correction.
The critical value was determined as $\sim 0.2411$ \cite{ON} by using
the fact that the lowest singlet and triplet excited levels cross at
this point in the thermodynamic limit.
For $\alpha_{\rm c}\le\alpha\le0.5$ and $\delta=0$ an excitation gap
opens as in Fig.\ 11(b), and the ground states are singlet and twofold
degenerate in the thermodynamic limit.
But this degeneracy is lifted for finite $N$ and $\alpha\ne 0.5$.
\par

At $\alpha=0.5$ the degenerate exact ground states are
known\cite{MG,SS} as:
$\Psi_\pm=\nu_\pm(\psi_1\pm\psi_2)$, where
\begin{eqnarray}
\psi_1&=&[\ 1,2\ ][\ 3,4\ ]\cdots[N-1,N],\nonumber \\
\psi_2&=&[\ 2,3\ ][\ 4,5\ ]\cdots[N,1],
\end{eqnarray}
and
$$
\nu_\pm=1\left/\sqrt{2^{N/2+1}\pm(-1)^{N/2}4}\right..
$$
Here, the square bracket indicates the singlet pair spin function:
\begin{equation}
[\ i,j\ ]=\alpha(i)\beta(j)-\beta(i)\alpha(j).
\end{equation}
$\Psi_+$ ($\Psi_-$) belongs to $Q=0$ ($\pi$).\cite{notetrans}
The spin excitations for the MG point were studied by introducing an
variational wave function,\cite{SS} in which an isolated doublet
(or spinon) pair plays an important role.
\par

For other values of $\alpha$ in the $J$-$J'$ model eq.\ (1.1),
the ground-state and thermodynamic properties are studied by the
exact diagonalization and a Quantum Monte Carlo method.\cite{TH}
\par

For a finite value of $\delta$, a finite gap always exists.
The exact ground state is known on the line $2\alpha+\delta=1$.\cite{SS}
In this case, $\psi_1$, in which singlet pairs sit on the bonds with
the larger exchange $J(1+\delta)$, becomes the unique ground state,
and $\psi_2$ (therefore $\Psi_\pm$) is no longer an eigenstate.
For $2\alpha+\delta>1$ the phase is ``spiral".
In this paper, we concentrate on the parameter range $2\alpha+\delta\le 1$
($\alpha, \delta\ge 0$), and leave the spiral phase for future
studies.
\par

For this model, our main concern is the spin dynamical structure factor
at zero temperature:
\begin{equation}
S(q,\omega)=\sum_n |\langle\Psi_n|S_q^z|\Psi_0\rangle|^2
\delta\left(\omega-(E_n-E_0)\right),
\end{equation}
where $\hbar=1$, $\Psi_n$ ($E_n$) is the $n$-th eigenfunction
(eigenvalue) of the system, and the ground state is indicated by $n=0$.
$n$ runs over all the eigenstates.
$S(q,\omega)$ is a quantity proportional to the spin contribution
in the intensity of an inelastic neutron scattering spectrum.
Since here the ground state is always singlet for finite systems,
only triplet states contribute to $S(q,\omega)$ due to the selection
rule.
For finite systems, $S(q,\omega)$ can be estimated by the recursion
(or continued-fraction) method.\cite{GB}
To this end, we expand the corresponding Green's function $G(z)$ as
\begin{equation}
G(z)=\langle \Psi_0|S_{-q}^z\frac{1}{z-{\cal H}}S_q^z|\Psi_0\rangle
    =S(q)C(z),
\end{equation}
where $S(q)=\langle\Psi_0|S_{-q}^zS_q^z|\Psi_0\rangle$ and
\begin{equation}
C(z)=\frac{1}{\displaystyle z-a_0-
\frac{\displaystyle b_1^2}{\displaystyle z-a_1-
\frac{\displaystyle b_2^2}{\displaystyle z-a_2\cdots}}}.
\end{equation}
For finite systems, $C(z)$ is comprised of a finite number of
$\delta$ functions.
Given the exact ground-state wave function, which we usually obtain
by the exact diagonalization, we can determine the coefficients $a_i$'s
and $b_i$'s through a Lanczos-like procedure.
Using this expansion, we can rewrite eq.\ (2.4) as a sum of residues,
\begin{eqnarray}
S(q,\omega)&=&-\frac{1}{\pi}\lim_{\eta\rightarrow 0}
                       {\rm Im}G(\omega+i\eta-E_0) \nonumber \\
  &=&S(q){\rm Re}\sum_\ell{\rm Res}\Bigl[C(\omega+i\eta-E_0),E_\ell\Bigr],
\qquad
\end{eqnarray}
where $\ell$ runs over all the poles of $C$.
\par

Here, we write down some technical points.
Actually, we calculate the contribution to $S(q,\omega)$ by
integrating $C$ for each pole with finite $\eta$,
typically 1-$5\times 10^{-5}J$.
Since the sum rule
\begin{equation}
S(q)=\int_0^\infty d\omega\ S(q,\omega),
\end{equation}
holds, the total contribution of residues for a fixed value of $q$ is
normalized to unity irrespective of the value of $\eta$.
We have checked that the total contribution in our calculation
is always $\gsim 0.9999$.
Next, to raise precision, it is important to start with a
ground-state vector as precise as possible.
Otherwise, numerical errors are rapidly mingled in the coefficients.
In fact, $a_i$'s and $b_i$'s accurate enough to be used seem to be
limited to $i\lsim20$-$30$ at most, so that we cut off higher coefficients
than $i=30$ mostly in this study.
This hardly affects poles and residues in low energy and with strong
intensity, but causes inaccuracy to some extent of the ones in high energy
and with weak intensity.
In this context, however, the results are justified for most cases
by confirming the regularity of the dependence on system size.
\par

Finally, we mention the finite-size scaling.
For the $J$-$J'$ model, there is no simple scaling function
which fits finite-size data correctly.\cite{ON}
However, by searching for a pertinent function case by case,
one can fulfill an extrapolation reliable enough to discuss
the physics.
We use the usual polynomial fit of order smaller than 10th, in addition to
the formula for $\delta\ne0$:\cite{Bouzerar}
\begin{equation}
\Gamma(N)=\Gamma(\infty)+\frac{\gamma}{N^m}
\exp\left(-\frac{N}{N_0}\right),
\end{equation}
where $\gamma$ and $N_0$ are fitting parameters, and $m=2$ (1)
is used if $\Gamma$ is energy (a quantity of energy difference
like the gap $\Delta$).
\par

\setcounter{equation}{0}
\section{Spin Dynamics of the $J$-$J'$ Model}
In this section, we focus on the $J$-$J'$ model eq.\ (1.1) without
coupling alternation.
By contrasting these results with the finite-$\delta$ cases, the
effect of $\alpha$ is definitely discriminated from that of $\delta$.
\par
\subsection{Excitation spectrum of the Heisenberg model}

Although the spin excitation of the 1D Heisenberg model has been
studied for a long period, it was quite recently that the accurate
behavior of the spectrum was clarified.
In contradiction to the spin-wave picture (spin 1),
the excitation spectrum of the Heisenberg model consists of
even number of free-spinon (spin 1/2) excitations.\cite{FT}
As the lowest-order contribution, two-spinon excitations form
a dominant continuum, the boundaries of which coincide with
the known curves:\cite{dCP}
\begin{equation}
\omega_{\rm \ell}(q)=\frac{\pi J}{2}\sin q,\quad{\rm and}\quad
\omega_{\rm u}(q)=\pi J\sin\frac{q}{2}.
\end{equation}
We call this continuum as the two-spinon continuum (TSC) in this
paper.
Some years ago, an approximate formula of $S(q,\omega)$ for the TSC
was proposed by M\"uller \etal from the consideration of the exactly
soluble $XY$ model, finite-size results and various sum rules,
as:\cite{Mueller}
\begin{equation}
S(q,\omega)=\frac{A}{\sqrt{\omega^2-\omega_\ell^2}}
\Theta(\omega-\omega_\ell)\Theta(\omega_{\rm u}-\omega).
\end{equation}
Here, $A$ is a constant of order 1, and $\Theta$ is the step function.
Actually, the intensity of this spectrum was roughly reproduced
by the neutron scattering experiments of KCuF$_3$,\cite{Tennant}
a good 1D antiferromagnet.
On the other hand, they simultaneously pointed out that it deviates
slightly from the exact one mainly due to higher-energy excitations
outside the TSC.
\par

Recently, the two-spinon and higher-order-spinon (HOS) contributions
to $S(q,\omega)$ have been exactly calculated by using formulae of
quantum group,\cite{BB} and compared with the exact diagonalization
as well as the above approximation.\cite{B}
According to these studies, although the formula eq.\ (3.2)
approximately represents the exact two-spinon spectrum,
eq.\ (3.2) misses the behavior in the vicinity of both boundaries.
Especially, it overestimates near the upper boundary,
under which there has to be a square root singularity in $S(q,\omega)$,
instead of a jump given by eq.\ (3.2).
Meanwhile, the two-spinon contribution to $S(q,\omega)$ occupies
72.89\% of the total weight; the HOS processes have not a little
effect on $S(q,\omega)$ as expected before by M\"uller \etal
\par

Here, we summarize the characteristics of the spectrum of the
Heisenberg model as a reference for later discussions, based on
our calculations by the recursion method.
We note that there have been a number of studies in this line
for the Heisenberg model.
In Fig.\ 1 $S(q,\omega)$ of the Heisenberg model is shown for $N=26$.
\par

Poles by the two-spinon processes have dominant intensity and are
situated on a series of sinusoidal curves in the TSC.\cite{Mueller}
The characteristics of the TSC are {\bf [a]} the intensity becomes strong
as $q$ approaches $\pi$ due to the logarithmic divergence of $S(q)$
at $q=\pi$.
{\bf [b]} At every fixed value of $q$, $S(q,\omega)$ is monotonically
decreasing function of $\omega$.
\par

On the other hand, the other poles are of the HOS contribution,
which continuously spread inside ($q/\pi\gsim 0.4$) as well as
outside of the TSC.
The existence of this contribution was suggested\cite{HB,Mueller}
in relation to the single-mode
approximation as discussed in Appendix A.
The intensity of these poles is considerably weak in comparison
with the two-spinon contribution, but severely dependent on $N$
and abruptly increases, as seen in Fig.\ 3(a).
The lower boundary of this HOS continuum is indicated by dashed
lines; it seems the lowest mode switches a couple of times,
and this boundary lowers with increasing $N$.
Incidentally, this curve looks similar to that of the lowest
quintet ($S=2$) excited levels (solid diamond for $N=14$), which are
described by four-spinon processes.\cite{BonnerB}
The four-spinon continuums of the triplet states and of the quintet
states may coincide in the thermodynamic limit, just as the TSC's
of the triplet states and of the singlet ones do.\cite{BonnerB}
At any rate, another important feature in the spectrum of the
Heisenberg model is {\bf [c]} there is non-two-spinon contribution,
although its intensity is weak.

\par
In the remainder of this subsection, we consider the dependence
on system size of the poles and their residues to discuss
whether a pole belongs to a continuum or forms an isolated branch.
Such a finite-size analysis was successfully used for the detection
of an isolated branch in the $S=1$ Heisenberg model,\cite{Takahashi}
although we cannot always draw a definite conclusion from this
analysis.
\par

Following this work, we plot residues of the poles belonging to
the lowest two two-spinon branches in Fig.\ 2(a).
Since both branches belong to a continuum, the value of
each residue ought to vanish in the thermodynamic limit.
As for the branch 1', the weight decreases with increasing
$N$.\cite{Takahashi}
However, the weight of the branch 2' is almost independent of $N$
for large $q$ and rather increases with increasing $N$ for medium $q$
within these system sizes.
In fact, we have checked various cases and found that the dependence
on $N$ of residues is sometimes not monotonic.
\par

On the other hand, Fig.\ 2(b) shows the dependence on $N$ of the
positions of poles for the lowest four two-spinon branches.
Although the branch 1' is scarcely dependent on $N$, other branches
have considerable dependence on $N$.
In the inset of Fig.\ 2(b) the pole positions at $q=\pi$ are plotted
versus $1/N^2$.
All the branches monotonically decrease as the system becomes large,
and seem to gather at $\omega=0$.
Note that $S(q,\omega)$ diverges as $1/\omega$ at $q=\pi$.
\par

As seen in this case, a pole which belongs to a continuum tends
to have appreciable dependence on system size at least either of its
position or of its residue.
On the other hand, if a pole has little dependence on $N$ both
of its position and of its residue, and unless adjacent poles converges
to its position, that pole probably forms an isolated branch.
However, as will be discussed in \S4.3, this will not be a necessary
condition for an isolated branch.
Thus, one should check carefully the dependence on $N$ of the
positions and residues in each case.
\par

\subsection{Excitation spectrum near the critical point $\alpha_{\rm c}$}

As mentioned in \S2, at $\alpha_{\rm c}=0.2411$ the system
eq.\ (1.1) is conformal invariant.
This fact reminds us of another conformal invariant model, the HS
model:\cite{HS}
\begin{equation}
{\cal H}=J\sum_{i<j}\frac{1}{d(|r_i-r_j|)^2}{\mib S}_i\cdot{\mib S}_j,
\end{equation}
where
$d(r)=(N/\pi)\sin(\pi r/N)$.
This model is exactly soluble, and not only ground-state but
thermodynamic\cite{H} and dynamical\cite{HZ} properties have been known.
$S(q,\omega)$ consists only of the two-spinon processes in contrast
to the Heisenberg model, and is given by a formula analogous to eq.\ (3.2),
as:
\begin{equation}
S(q,\omega)=\frac{1}{4}
\frac{\Theta(\omega-\tilde\omega_{\ell+})\Theta(\omega-\tilde\omega_{\ell-})
\Theta(\tilde\omega_{\rm u}-\omega)}
{\sqrt{(\omega-\tilde\omega_{\ell+})(\omega-\tilde\omega_{\ell-})}},
\end{equation}
where the boundaries of the continuum are given by
$\tilde\omega_{\ell-}(q)=Jq(\pi-q)/2$,\
$\tilde\omega_{\ell+}(q)=J(q-\pi)(2\pi-q)/2$
and $\tilde\omega_{\rm u}(q)=Jq(2\pi-q)/4$.
\par

With these in mind, let us look at the case of
$\alpha\lsim\alpha_{\rm c}$.
Even if $\alpha$ is switched on, $S(q,\omega)$ basically keeps the
feature of the Heisenberg model within the gapless regime.
As for the HOS processes, however, their total contribution considerably
decreases from the value of the Heisenberg model as shown in Fig.\ 3(a),
which shows the sum of residues by the HOS processes for small values
of $q/\pi$.
In Fig.\ 3(b), the maximum values of the above sum between
$0\le q/\pi\le 0.5$ are plotted as a function of $\alpha$ for $N=26$.
For $\alpha\sim 0.2$ the HOS contribution becomes minimal.
\par

In view of the conformal field theory, namely as for low-energy
processes, these two conformal invariant models are related
to each other as follows.
If we cut off the exchange coupling longer than the second neighbor
in the HS model, the residual part becomes the $J$-$J'$ model with
$\alpha=0.25$.
The marginally irrelevant operators in the discarded long range
part cancel out with the corresponding ones in the second neighbor
coupling of $0.25-\alpha_{\rm c}$.
Thus, it is natural to expect that the behavior of $S(q,\omega)$
for small $\omega$ at $\alpha\sim\alpha_{\rm c}$ resembles that of
the HS model.
Strictly speaking, however, the HOS contribution never vanishes
and the minimum contribution is not at $\alpha_{\rm c}$ but
$\sim 0.18$-0.2.
This means that the behavior of higher-energy processes deviates
to some extent from the expectation in the infrared limit.
\par

Such tendency can be seen in Fig.\ 4, which shows $S(q,\omega)$
for $\alpha=0.2$.
The shape and the size of the main continuum almost coincide with those
of the HS model; at the upper edge of the continuum, $S(q,\omega)$
is more likely to have a jump than that in Fig.\ 1.
For more quantitative discussions on $S(q,\omega)$ the introduction
of the $n$-th frequency moment defined as,
\begin{equation}
K^{(n)}(q)=\int_0^\infty d\omega\ \omega^n S(q,\omega).
\end{equation}
is convenient.
First, we consider static susceptibility $\chi(q)=K^{(-1)}(q)$,
which is depicted for some values of $\alpha$ in Figs.\ 5(a)
and 5(b).
\par

For the Heisenberg model, the uniform susceptibility $\chi(0)$
is given by the Bethe Ansatz;\cite{Griffiths} we choose the
normalization as $\chi(0)J=1/(2\pi^2)$.
For a general value of $q$, diagonalization results have been discussed
recently;\cite{B} $\chi(q)$ is expected to diverges as $1/(\pi-q)$
with some logarithmic correction.
What we should add here is that as in the inset of Fig.\ 5(a) slight
dependence on $N$ is observed for small $q$, and that the extrapolated
value for $q=0$ seems to deviate subtly upward from Griffiths' value.
This may possibly means the existence of some logarithmic singularity
in the small-$q$ limit owing to a non-fixed-point model.
\par

As $\alpha$ increases, $\chi(q)$ increases for medium values of $q$.
The tendency becomes opposite for the higher-order frequency moments,
as will be mentioned shortly.
As discussed above, the curve for $\alpha=0.2$, rather than
$\alpha_{\rm c}$, coincides well with that of the HS model
except for $q\sim 0$, where the long-range part of the coupling
probably plays a major role.
By the way, $\chi(q)$ for the HS model is obtained by integrating
eq.\ (3.4) as,
\begin{equation}
\tilde\chi(q)J=\frac{1}{2}
\frac{1}{\sqrt{|\tilde\omega_{\ell-}\tilde\omega_{\ell+}|}}
\left[\arcsin\left(\frac{q}{\pi}-1\right)+\frac{\pi}{2}\right].
\end{equation}
In contrast with the Heisenberg case, the leading power is linear
in $q$ for $q\rightarrow 0$ and $\tilde\chi(q)\propto 1/(\pi-q)$
without logarithmic correction for $q\rightarrow\pi$.\cite{Arikawa}
\par

We discuss the cases of $\alpha>\alpha_{\rm c}$ together here.
Since the system has a finite gap, $\chi(q)$ vanishes for
$q\rightarrow 0$.
As a sign of it, the abrupt decrease of $\chi(q)$ can be seen
for large values of $\alpha$.
The value of $q$ at which $\chi(q)$ drops corresponds to $\pi/\xi$,
where $\xi$ is the spin correlation length.
For the MG case ($\alpha=0.5$), the spin correlation $S_i$ vanishes
for $i\ge 2$; the anomaly exists at $q\sim\pi$.
By extrapolation we obtain $\chi(\pi)J\sim0.843$.
\par

Leaving the discussion about $K^{(0)}(q)=S(q)$ and $K^{(1)}(q)$
for \S3.3 and Appendix A respectively, here we consider $K^{(2)}(q)$
and $K^{(3)}(q)$.
As the order $n$ of the frequency moment becomes higher, the effect
of high-energy processes is amplified.
In Figs.\ 6(a) and 6(b), the second and third frequency moments are
plotted, respectively.
As $n$ increases, the values of the $J$-$J'$ model for $q\sim\pi$
are enhanced in comparison with that of the HS model given by
the integration of eq.\ (3.5):
\begin{eqnarray}
&&\frac{\tilde K^{(2)}(q)}{J^2}=
-\frac{q}{128}(2\pi-q)(7q^2-14\pi q+6\pi^2) \nonumber \\
&&\qquad-\frac{1}{32}(q-\pi)^2(2q^2-4\pi q+3\pi^2)
\ln\left(1-\frac{q}{\pi}\right), \\
&&\frac{\tilde K^{(3)}(q)}{J^3}=\frac{q}{768}(2\pi-q) \nonumber \\
&&\qquad\times(28q^4-112\pi q^3+169\pi^2q^2-114\pi^3q+30\pi^4)
\qquad\nonumber\\
&&\qquad+\frac{1}{64}(q-\pi)^4(2q^2-4\pi q+5\pi^2)
\ln\left(1-\frac{q}{\pi}\right).\qquad
\end{eqnarray}
A main cause is probably the HOS contribution of the $J$-$J'$ model
outside the TSC, as pointed out for the Heisenberg case.\cite{B}
The enhancement for small $q$ and large $\alpha$ is owing to the
considerable intensity beyond the upper edge of the main continuum
in the gap regime as in Fig.\ 3(b).
This is one of the topics in \S3.3.
\par

Finally, we point out one more important fact that the height of the
TSC becomes shorter monotonically with increasing $\alpha$.
The highest point of the TSC continuum (at $q=\pi$) is $\pi J$ for
the Heisenberg model, and $\pi^2J/4$ for the HS model, which is
comparable with the case of $\alpha\sim\alpha_{\rm c}$.
The influence of it appears also in Figs.\ 6(a) and 6(b); as $\alpha$
increases, $K^{(n)}(q)$ for large $n$ generally decreases.
We return to this point later when we estimate the value of $J$ by
experimental spectra.
\par

Having described mainly the resemblance of $S(q,\omega)$ for
$\alpha\sim\alpha_{\rm c}$ to that of the HS model, we close this
subsection by summarizing the properties of $S(q,\omega)$ for
$\alpha\lsim\alpha_{\rm c}$ (for example, Fig.\ 4) for later discussions.
Basically, the behavior of $S(q,\omega)$ remains having common features
with the Heisenberg model; we emphasize repeatedly
{\bf [a]} the intensity of $S(q,\omega)$ becomes considerably strong
as $q$ approaches $\pi$ due to the divergence of $S(q)$.
{\bf [b]} The intensity of $S(q,\omega)$ is strongest at the
lower edge and a monotonically decreasing function of $\omega$ for
every $q$.
As for the point [c], although non-two-spinon contribution exists,
their intensity becomes extremely weak for $\alpha\sim\alpha_{\rm c}$,
where the features of $S(q,\omega)$ resemble that of the HS model.
\par
\subsection{Excitation spectrum for $\alpha>\alpha_{\rm c}$}

Since the characteristics of the gap regime are typically realized
in the MG case ($\alpha=0.5$), we take it up first in this subsection.
In this case we use $\Psi_\pm$ as the starting vectors in the recursion
method.
In Fig.\ 7(a) we show the residues of $C(z)$ [eq.\ (2.6)] for
$N=24$; we plot both contributions from $\Psi_+$ ($Q=0$)
and $\Psi_-$ ($Q=\pi$) together, therefore the spectrum is completely
symmetric with respect to $q=\pi/2$.\cite{notereliability}
In Fig.\ 7(b) $S(q,\omega)$ is shown for $N=26$.
We itemize their characteristics in comparison with the gapless
regime below.
\par

{\bf [1]} There is appreciable intensity outside the main continuum
not only for the small-$q$ region [see also Fig.\ 3(b)] but in the
vicinity of $q=\pi$.
On the contrary, there is little high-energy contribution
around $q=\pi/2$.
\par

{\bf [2]} $S(q,\omega)$ is not a monotonically decreasing function
of $\omega$; the upper edge of the continuum for $q>\pi/2$ has stronger
intensity than the inner area of the continuum.
\par

{\bf [3]} The region near the lower edge of the continuum has strong
weight, like the gapless cases.
However, the weight of $S(q,\omega)$ between $q=\pi/2$ and $\pi$ is
comparable, because the residues become huge around $q=\pi/2$,
as well as $S(q)$ does not diverge at $q=\pi$.
\par

{\bf [4]} An isolated branch appears for $0.45\lsim q/\pi\lsim 0.55$.
As seen in Fig.\ 7(a) this isolated branch becomes a pure single
mode for $q=\pi/2$ with $\omega=J$ independent of $N$.
The weak intensity in high energy and the strong intensity at the
lower edge around $q=\pi/2$ mentioned in [1] and [3] respectively
are attributed to this isolated branch.
\par

These features are in sharp contrast with the ones [a]-[c] for
$\alpha<\alpha_{\rm c}$.
\par

Let us look at the spectra more closely.
The structure of the continuum is more complicated than the TSC
in the gapless regime; near the upper edge it is not easy
to trace each branch, while a series of dominant branches can
be distinguished in the lower half of the continuum.
In particular, near the lower edge, a number of branches with
comparable weight seem to gather.
The lowest branch does not have the strongest intensity.
\par

To check the dependence on $N$, we show the residues of the four
lowest dominant branches in Figs.\ 8(a) and 8(b).
For $q/\pi\gsim 0.55$ the weight of every branch depends severely
on system size.
Although the branches 1' and 2' monotonically decrease and the branch 4'
monotonically increases with increasing $N$, the branch 3' shows
complicated behavior.
Positions of the corresponding poles are shown in Fig.\ 8(c).
The higher the energy of the branch is, the severer the dependence is.
As shown in the inset of Fig.\ 8(c), which shows the pole positions
versus $1/N^2$ at $q=\pi$, every branch seems to converge to the lower
edge.
\par

It is clear that the spectrum forms a continuum for $q/\pi\gsim 0.55$,
because both the residues and the poles are considerably dependent on
system size.
On the other hand, as in the Figs.\ 8(a) and 8(b), for $q/\pi\lsim 0.55$
the weight of the branches 1' and 2' is enhanced rapidly as $q$
approaches $\pi/2$ without dependence on $N$.
Inversely, the branches 3' and 4' seem to vanish.
As for the pole positions of the branches 1' and 2', there is
little dependence on $N$ [Fig.\ 8(c)].
Consequently, according to the discussion in \S3.1, an isolated branch
with weight more than 90\% seems to occur in this region.
In fact, at $q=\pi/2$ it is proven that the spectrum becomes a pure
single mode as shown in Appendix B.
Incidentally, a similar spectrum was obtained by using a variation
method;\cite{SS} in addition to a continuum, a bound state below the
continuum was derived for $0.36\lsim q/\pi\lsim 0.64$.
\par

Next, we touch on the point [2].
In Fig.\ 9 the residues at $q=\pi$ are plotted versus $\omega$.
Near the lower edge, there exists a large peak like the gapless cases.
However, it is not clear in this case whether this peak diverges, and
whether it is situated exactly at the lower edge.
Anyway, the weight in the continuum once has a broad minimum
($\omega/J\sim 1.5$-1.8), then another peak appears near the upper
edge ($\omega/J\sim 2$).
This peak seems less dependent on $N$, and has tails in both
sides of $\omega$.
Similar tendency can be seen for other values of $q$.
This profile is in sharp contrast with those in the gapless region.
\par

Having focused on the MG case, here we mention the cases of smaller
values of $\alpha$.
In these cases, the ground state is not twofold degenerate except
for the thermodynamic limit, thus we treat excitations only
from the nondegenerate ground state for finite $N$.
In the thermodynamic limit, the two ground states with $Q=0$ and
$\pi$ will give contributions mutually symmetric with respect
to $q=\pi/2$ just like the MG case.
\par

Figures\ 10(a)-10(c) show $S(q,\omega)$ for $N=26$ and $\alpha=0.35$,
0.4 and 0.45, respectively.
In these figures we can recognize that the characteristics of the
MG model [1]-[4] still remain.
Comparing these with Fig.\ 4 ($\alpha=0.2$) and Fig.\ 7(b) ($\alpha=0.5$),
we find that the spectrum gradually changes from the gapless feature
to the one of the MG case.
\par

Finally, for later discussions, we represent some static quantities;
Fig.\ 11(a) shows the ground-state energy, and Fig.\ 11(b) the lowest
excitation at $q=\pi/2$ and the gap $\Delta$.
These values are extrapolated from the finite-size data by the polynomial
fit of second and fourth orders.
There is no critical behavior at $\alpha=\alpha_{\rm c}$.
\par

The static structure factor $S(q)$ both for the Heisenberg model\cite{B}
and the HS model\cite{GV,HZ} is logarithmically divergent at $q=\pi$.
Accordingly, the situation is basically the same for the gapless regime
of the $J$-$J'$ model.
On the other hand, as $\alpha$ increases beyond $\alpha_{\rm c}$,
the divergence at $q=\pi$ is suppressed.
For the MG point $S(q)=(1-\cos q)/4$, namely $S(q)$ is not only
convergent but upward convex near $q=\pi$.
In Fig.\ 12 $S(q)$ for large $\alpha$ is shown.
Even for $\alpha=0.4$, $S(q)$ is considerably enhanced near $q=\pi$.
Furthermore, it increases with increasing system size.
On the other hand, for $\alpha=0.45$ the dependence on $N$ of $S(\pi)$
is in the direction of suppression.
\par

Anyway, concerning the excitation in the regime of
$\alpha>\alpha_{\rm c}$, we are still far from understanding
in the standard of the Heisenberg model.
Further studies are necessary.
\par

\section{Comparison with Spin-Peierls Compounds}

Based on the discussion in the preceding section, we consider
various aspects of the SP transition in CuGeO$_3$ and
$\alpha'$-NaV$_2$O$_5$.
\par

\subsection{Complete spectrum of inelastic neutron scattering}

Recently, a complete spin excitation spectrum of CuGeO$_3$ was
obtained by the inelastic neutron scattering by Arai \etalp\cite{Arai}
The gap and the size of the continuum in this spectrum are consistent
with the previous neutron experiments.\cite{Nishi,Hirota1,Regnault}
At a glance on the spectrum for $T=10$K (Fig.\ 1 of ref.\ 19),
it looks like the TSC of the Heisenberg model.
However, as the authors pointed out, there are points which
are incompatible with the Heisenberg model.
We itemize such points below including the authors' words.
\par
{\bf [0']} The shape of the lower and the upper edges of the continuum
is wedge-like and does not have the roundness characteristic of
sine ($\alpha=0$) and quadratic ($\alpha\sim\alpha_{\rm c}$) curves.
Concerning this point, compare Figs.\ 18(a) and 19(a), for instance.
\par
{\bf [1']} There is appreciable intensity above the continuum,
especially for $q\sim \pi$.
\par
{\bf [2']} There is ``rampart" or ridge of scattering surrounding
a valley in the spin continuum.
\par
{\bf [3']} There exists a peak in intensity on the lower boundary
of the continuum at zone boundaries (0,0,$\pi/2$) and
(0,0,$3\pi/2$).
\par

The points [1']-[3'] completely correspond to what we have summarized
as [1]-[3] for $\alpha>\alpha_{\rm c}$ in \S3.3, respectively.
In this viewpoint, let us compare Fig.\ 1 of ref.\ 19 with
Figs.\ 7(b) and 10(a)-10(c) in this paper, before taking account of
the effect of coupling alternation.
Among the above four panels Fig.\ 10(c) ($\alpha=0.45$) seems to bear
the best resemblance.
\par

As for the point [1'], the intensity of every pole over the upper
edge of the main continuum ($q\sim \pi$) in Fig.\ 10(c) corresponds
to the one in Fig.\ 1 of ref.\ 19.
For example, a horn-like short branch jutting out upward from
$q=\pi$ just above the main continuum ($\omega/J\sim 2.1$) is distinctly
reproduced.
Incidentally, because the spectrum should be symmetric with respect
to $q=\pi$, the asymmetric intensity in the experimental spectrum,
especially for $q>\pi$, is due to the contamination by phonons etc.,
as the authors mentioned.
\par

Concerning the point [2'], the strong intensity at the upper edge
and the weak intensity in the upper half of the continuum
for $q\sim2\pi/3$\ -\ $\pi$ are quite similar between the two figures.
Recall the profile in Fig.\ 9 for the MG case.
\par

As for the point [3'], the experimental intensity near the lower
edge is almost the same between $q=\pi/2$ and $\pi$.
Furthermore, the maximal points seem to be situated slightly above
the exact lower edge, which we have suggested in \S3.3, although
it is not clear whether this is due to its intrinsic property
or thermal broadening.
\par

For smaller values of $\alpha$ [Figs.\ 10(a) and 10(b)], the tendencies
deviate from the experimental spectrum.
For example, in Fig.\ 10(a) for $\alpha=0.35$, the strongest intensity,
as to the lower edge, is clearly at $q=\pi$.
If $\alpha\sim\alpha_{\rm c}$, the intensity over the continuum
considerably attenuates, as discussed in \S3.2.
Consequently, the value of $\alpha$ should be fairly larger than
$\alpha_{\rm c}$.
\par

More quantitative fit is possible; in Fig.\ 13 we compare integrated
amplitude $S(q)$ (Fig.\ 12) with the neutron data, which are taken
from Fig.\ 2 of ref.\ 19.
In the experimental results, there is neither divergence nor abrupt
increase for $q\sim\pi$ characteristic of smaller values of $\alpha$.
It reads $\alpha\sim 0.45$ for $T$=10K and $\alpha\sim 0.5$ for
$T=20$K ($T_{\rm SP}=14$K), which is consistent with the above direct
comparison of $S(q,\omega)$.
This indicates that the value of $\alpha$ hardly changes through
$T_{\rm SP}$, or reduces slightly for $T<T_{\rm SP}$ if any.
Incidentally, there is a tendency of the experimental data to be
rather smaller than the diagonalization result for $q\sim\pi$.
This is probably due to two facts.
One is that in obtaining $S(q)$ by integrating experimental
$S(q,\omega)$, the higher-frequency contribution than 32meV (the top
of the continuum) is cut off to exclude the contribution by phonon.
Actually, there remains magnetic contribution of several percent,
as mentioned above.
The other is that the introduction of $\delta$ somewhat reduces
$S(q)$ at $q=\pi$.
In this connection, if a smaller value of $\alpha$ is assumed,
a considerable large value of $\delta$ is needed to reproduce
the experiment, for example, $\delta\gsim 0.03$ even for $\alpha=0.4$;
such a parameter set is incompatible with the value of $\Delta$,
as we will see later.
\par

Thus, we have found that most features of this neutron spectrum
are explained by the $J$-$J'$ model with $\alpha=0.4$-0.45, even
without introducing $\delta$.
\par

Next, let us check whether or not the above aspects are also
realized by introducing $\delta$ without $\alpha$ or with smaller
values of $\alpha$.
Figures 14(a)-14(c) show $S(q,\omega)$'s for $\delta=0.05$
and three values of $\alpha$.\cite{notebz}
To emphasize the effect of $\delta$, a fairly large value of $\delta$
is used for CuGeO$_3$ ($\delta$ is of order 0.001).
In these figures some new features can be seen.
For example, as will be discussed in \S4.3, there appears an opening
of intensity between the lowest branch and the above continuum,
especially for large $\alpha$.
However, for the case of $\alpha<\alpha_{\rm c}$ [Figs.\ 14(a) and 14(b)]
the characteristics [1]-[3] for $\alpha>\alpha_{\rm c}$ discussed
in \S3.3 do not appear even if $\delta$ is added, while these
characteristics are preserved for $\alpha>\alpha_{\rm c}$ if $\delta$
is added, as in Fig.\ 14(c).
In other words, the characteristics [1]-[3] appear only
when $\alpha>\alpha_{\rm c}$.\cite{notebigd}
Furthermore, these characteristics do not become manifest unless
$\alpha$ is fairly larger than $\alpha_{\rm c}$.
For smaller values of $\delta$, the characteristics [1]-[3] do not
appear for $\alpha<\alpha_{\rm c}$, of course.
Thus, these facts exclude at least the possibility of
$\alpha<\alpha_{\rm c}$, which is contrary to the discussion
of Castilla \etalp\cite{Castilla}
We will point out a mistake in their assumption in \S4.3.
\par

\subsection{Estimation of $J$ and $\delta$}

First, we discuss the value of $J$.
The first estimation of $J$ was given by a neutron experiment as
$J\sim120$K,\cite{Nishi} in which the maximal point of the lowest
branch in the spectrum was compared with $J\pi/2$, which is obtained
from the des Cloizeaux-Pearson curve [the first equation of eq.\ (3.1)]
for the Heisenberg model.
However, the experimental spectrum has to be compared with that
of eq.\ (2.1).
As we have already shown in Fig.\ 11(b), the size of the continuum
decreases with increasing $\alpha$.
The maximal value of the lower edge ($q=\pi/2$) decreases from
$J\pi/2$ to $J$, as $\alpha$ changes from 0 to 0.5.
Since $\alpha=0.4$-0.45, $J$ is modified into a larger value by
$\sim$50\%; thus we have $J\sim 180$K.
\par

In the above estimation of $J$ we have neglected the effect of
$\delta$; let us check it below.
In Fig.\ 15 the maximal values of the lower edge are plotted
as a function of $\delta$ for some values of $\alpha$.
In contrast with the effect of $\alpha$, the spectrum size weakly
depends on $\delta$.
Although it abruptly increases for small $\delta$ and large $\alpha$,
the increment is at largest a few percent, because $\delta$ is
extremely small.
Consequently, the coupling alternation scarcely affects the
estimation of $J$.
\par

Now, we refer to the estimations of $J$ by other means.
Many authors tried to evaluate $\alpha$ and $J$ by comparing with
$\chi(T)$ by experiment,\cite{Hase} which shows a broad peak
around 56K.
Among them, Riera and Dobry obtained $\alpha=0.36$ and $J=160$K,
by fitting the maximal point of $\chi(T)$.\cite{Riera}
Since this fit was quite good for $T>56$K, following studies
often adopted these values.
\par

Using Faraday rotation, Nojiri \etal\cite{Nojiri} estimated $J$ at
183K; they compared the field at which magnetization was saturated with
the formula for the Heisenberg model: $g\mu_{\rm B}H_{\rm s}=2J$.
Although in this case the corresponding value for eq.\ (2.1) should
be used for accuracy, it is convenient to use, instead, the known formula
for the $J$-$J'$ model:\cite{TH}
\begin{equation}
g\mu_{\rm B}H_{\rm s}/J=\left\{
\begin{array}{ll}
2 & 0\le\alpha\le 0.25 \\
1+2\alpha+1/8\alpha & \alpha\ge 0.25
\end{array}
\right.
.
\end{equation}
Unless the variation by $\delta$ is singular at $\delta=0$, this
estimation is not poor because $\delta$ is very small.
As a result, the value is slightly modified as $J=178$K, 173K, 168K
for $\alpha=0.35$, 0.4, 0.45, respectively.
\par

Thus, the estimation of $J$ by the neutron experiments, $J\sim 180$K,
is consistent with other estimations.
By using $\alpha=0.45$ (0.4) and $J=180$K, the gap energy without coupling
alternation is evaluated at 19K (5K), which must be a value too small
to make a gap by itself above $T_{\rm SP}$.
\par

Next, we estimate the coupling alternation parameter $\delta$.
The excitation gap was estimated by some experiments.
In particular, the neutron scattering experiment can directly observe
the gap as a function of temperature $\Delta(T)$, and concluded
$\Delta(0)=2.1$meV (24K),\cite{Nishi} which is consistent with
other experiments like NMR,\cite{Yasuoka} specific heat,\cite{Sahling}
magnetic susceptibility $\chi(T)$,\cite{Hase} etc.
\par

In Figs.\ 16(a) and 16(b), the gap $\Delta$ versus $\delta$ obtained
by the exact diagonalization for eq.\ (2.1) is depicted.
In Fig.\ 16(b), the dash-dotted (dashed) line corresponds to
$\Delta=24$K and $J=180$K (160K).
If we assume $\alpha=0.45$ (0.4) and $J=180$K, $\delta$ becomes
as tiny as $0.001$ (0.005).
This value is much smaller than the previous estimations:
$\delta=0.012$-0.014.\cite{Riera,Bouzerar}
This difference is originated in the fact that $\Delta$ is an
abruptly changing function of $\alpha$ for $\sim 0.5$ and of
$\delta$ for $\sim 0$.
\par

Experimentally, the atomic displacement due to the lattice dimerization
was measured for CuGeO$_3$ by neutron diffraction.\cite{Hirota1}
The displacement ratio $\delta_\ell$ is very small;
$\delta_\ell\sim 0.001$ both for Cu along the $c$ axis parallel to
the 1D chain and for O(2) along the axes perpendicular to the chain.
This value is too small to detect in some other means.\cite{NMR}
However, it is still unknown how this small displacement is connected
to a small exchange coupling.
\par

In CuGeO$_3$, CuO$_6$ octahedra link mutually with edge sharing
along the $c$ axis; the angle between Cu-O-Cu is 98$^\circ$.
The exchange coupling, which is determined by the overlap of the
involved localized orbitals, tends to be weakly ferromagnetic
in the case of the right angle, as Kanamori classified many years
ago.\cite{Kanamori}
On the other hand, recently Geertsma and Khomskii argued that
for the realization of antiferromagnetic coupling the side-group
effect of Ge plays an important role.\cite{Khomskii}
Thus, we cannot assert the relation $\delta_\ell\propto\delta$
necessarily holds.
\par

By the way, from the above discussion one can expect relatively
large long-range exchange couplings survive in edge-sharing materials.
Hence, we have checked the third-neighbor's effect for various
strength of $J^{(3)}$ ($>0$) and for some values of $\alpha$ and $\delta$.
The common aspect among the resultant spectra is that the top
of the lower edge of the continuum becomes broad and flat around
$q=\pi/2$ with the spectral size almost unchanged.
This aspect is quite different from the characteristics of
CuGeO$_3$.
Thus, the third-neighbor exchange is found negligible.
\par

Now, we refer to the second inorganic SP compound
$\alpha'$-NaV$_2$O$_5$,\cite{Ueda} in which VO$_5$ pyramids link
mutually with corner sharing in the 1D direction.
Since $\chi(T)$ for $T>T_{\rm SP}$ is well fitted by the Bonner-Fisher
curve\cite{BF} with $J=$560K, the frustration by $J'$ seems small
in this compound.
The gap is estimated at $\Delta=98K$ by NMR.\cite{Ohama}
Using these values or another estimation of $J=441$K and
$\Delta=85$K,\cite{Weiden} we have $\delta=0.043$-0.048 from
the curve of $\alpha=0$ in Fig.\ 16(a).\cite{Poil2}
Since the relation $\delta_\ell\propto\delta$ holds in this
case, the displacement due to dimerization is probably much larger.
In fact, a split in the Na-NMR spin-echo intensity corresponding
to the chain direction has been observed below $T_{\rm SP}$,\cite{Ohama}
in contrast with CuGeO$_3$.\cite{Yasuoka,NMR}
\par

Before closing this subsection, let us look at the SP transition
in the light of energy.
We depict the total magnetic energy $E/J$ for $\delta=0$ and
$\delta\ne 0$ in Figs.\ 11(a) and 17, respectively.
As temperature is lowered, the lattice will be distorted so as
to minimize $E/J$.
Thus, the displacement at $T=T_{\rm SP}$ will tend to enhance $\delta$
as well as $J$ and lower $\alpha$.
This tendency agrees with the one seen in Fig.\ 13, although
the thermal fluctuation may be also involved in this case.
\par

\subsection{Further discussions}

In this subsection, we will consider a couple of issues
in relation to the experiments.
\par

First, we take up $\Delta(T)$.
Harris \etal\cite{Harris} discussed this issue by combining three
materials:
(1) The dependence on temperature of the intensity of a superlattice
peak: $I(T)\propto [T_{\rm SP}-T]^{2\beta}$.
The work by neutron diffraction\cite{Hirota2,Martin} and X-ray
diffraction\cite{Harris2} concluded $\beta\sim 0.33$ in agreement.
(2) The spontaneous contraction along the $b$ axis with the
Ginzburg-Landau theory: $I\propto \delta_\ell^2$.
(3) The Cross-Fisher theory: $\Delta\propto\delta^{2/3}$.
Thus, they obtained $\Delta\propto [T_{\rm SP}-T]^{2a\beta}$ with
$a=1/3$, tacitly assuming $\delta\propto\delta_\ell$.
\par

Later, Castilla \etal\cite{Castilla} excluded the possibility of
$\alpha>\alpha_{\rm c}$, based on the above discussion and the
renormalization-group theory.
They assumed $\Delta\propto\delta^{2/3}$ with negligible logarithmic
correction throughout the gapless regime, while $\Delta\propto\delta$
for $\alpha>\alpha_{\rm c}$.
However, this assumption is false.
As in Fig.\ 16(a), $\Delta$ is nearly proportional to $\delta$ for
small $\delta$ and $\alpha=0$.
The leading power $\delta^{2/3}$ is smeared out due to the severe
logarithmic correction, even for large values of $\delta$.
This aspect had been already pointed out by the diagonalization
studies.\cite{Soos,BonnerB}
On the contrary, the leading power $\delta^{2/3}$ was clearly observed
for $\alpha=\alpha_{\rm c}$ in the density matrix renormalization group
calculations by Chitra \etal,\cite{Chitra} because the model becomes
conformal invariant and there is no logarithmic correction, as mentioned
in \S3.2.
Our data for $\alpha=0.2$ is consistently fitted with a power of
$\sim0.71$.
As $\alpha$ becomes larger, the seeming power further decreases
with a finite gap at $\delta=0$ and approaches 0.5 near the MG point.
These results are rather contrary to the assumption by Castilla \etalp
Consequently, there is no theoretical reason that $\alpha$
has to be smaller than $\alpha_{\rm c}$.
\par

Actually, the value $a$ in the power of $\Delta(T)$ was estimated
at $\sim 0.15$,\cite{Martin} which is much smaller than the seeming
power in Fig.\ 16 of $\sim 0.26$ for $\alpha=0.45$, as well as
the prediction of the Cross-Fisher theory of 1/3.
However, the discussion of power itself is meaningless for the gap,
which has serious logarithmic modification.\cite{notelog}
Furthermore, it is not certain whether the relation $\delta_\ell\propto\delta$
holds for edge-sharing materials like CuGeO$_3$, as discussed before.
On the other hand, $\Delta(T)$ for $\alpha'$-NaV$_2$O$_5$ also shows
a similar abrupt increase at $T=T_{\rm SP}$.\cite{Fujii}
These may indicate that $\Delta(T)$ is independent of the details of
the exchange coupling.
\par

Second, we consider the double gap reported in a recent neutron
experiment for CuGeO$_3$.\cite{Ain}
According to it, gap-like weak intensity was observed above the
lowest peak at 2.1meV for $q=\pi$; this ``second gap" had a similar
width with the first one.
\par

Let us return to Figs.\ 14(a)-14(c) for $\delta=0.05$.
Comparing these figures, we recognize that the aspects are
different between Figs.\ 14(a) [or 14(b)] and 14(c).
In the former case ($\alpha<\alpha_{\rm c}$) the lowest branch
for $\delta=0$ [Figs.\ 1 and 4]
splits into two and the opening between them becomes wider
with increasing $\alpha$; at $q=\pi$, however, these two branches
merge and the second gap seems to vanish.
Meanwhile, in the latter case ($\alpha>\alpha_{\rm c}$) the second
gap is open for all the range of $q$.
Thus, this issue is highly dependent, at least quantitatively,
on the value of $\alpha$ and $\delta$.
Here, we focus on the plausible parameters for CuGeO$_3$ and
$\alpha'$-NaV$_2$O$_5$, leaving more general discussions for
future publication.
\par

To begin with, we consider the case of $\alpha=0.45$, namely CuGeO$_3$.
Shown in Figs.\ 18(a) and 18(b) is the dependence on $N$ of the pole
position and the residue respectively corresponding to the two lowest
branches for $\delta=0.01$.
We use a somewhat larger, but sufficiently small, value of $\delta$
than our estimation so as to find its effect.
\par

First, we discuss the branch 1'.
Both the pole position and the residue of the branch 1' scarcely
depend on system size except for $q\sim 0$ or $\pi$.
For $q=\pi$, relatively reliable extrapolation for both quantities
is possible, as shown in the inset of Fig.\ 18(a) and by an arrow
in Fig.\ 18(b).
Moreover, since the pole position of the branch 2' is always away from the
branch 1', as will be discussed below, the branch 1' forms an
isolated mode.
\par

Next, as for the branch 2' both the pole position and the residue
are almost independent of system size for medium values of $q$,
although the behavior of the residue is complex.
For $q=\pi$, the pole position converges to a value which is higher
than that of the branch 1', as shown in the inset of Fig.\ 18(a).
On the other hand, we are not certain that the residue of the branch 2'
at $q=\pi$ has finite weight for $N\rightarrow\infty$.
Concerning the branch 3' (not shown in the figures), both the pole
position and the residue depend on $N$, but we cannot elicit
definite answers from our data whether its pole position converges
to that of the branch 2'.
Consequently, the branch 2' probably forms another isolated branch,
although there remains the possibility of the lower edge of the
continuum.
The branch 3' belongs to the continuum.
\par

Anyway, the opening between the branches 1' and 2' exists for
every value of $q$.
This opening is thin for medium values of $q$ and relatively
wide at $q=\pi$.
Nevertheless, the width of this second gap at $q=\pi$ is fairly smaller
than that of the first gap, as shown in the inset of Fig.\ 18(a).
Incidentally, if the branch 2' is an isolated branch at $q=\pi$ too,
the third gap will also exists.
\par

We have confirmed this situation remains the same for some similar
values of the parameters.
On the other hand, for larger values of $\delta$ (for example $\alpha=0.45$
and $\delta=0.1$), the branches 1' and 2' form clear isolated modes,
as studied formerly.\cite{Bouzerar}
Even in this case, however, the width of the second gap is found
still smaller than the first gap.
This result is in sharp contrast with an RPA calculation,\cite{Uhrig}
in which these two gaps have the same width.
\par

Now, we turn to the case of $\alpha=0$ and $\delta=0.05$ with
$\alpha'$-NaV$_2$O$_5$ in mind.\cite{Poil2}
In Fig.\ 19(a) we show the pole positions of the lowest three
branches with dominant intensity.
The branches 1' and 2' are almost independent of system size
for all the range of $q$.
A conspicuous point is that these two branches merge into a
single pole at $q=\pi$.
On the other hand, the poles belonging to the branch 3' have
severe dependence on system size.
Moreover, we find by the extrapolation of the poles at $q=\pi$
that the branch 3' converges to the branch 2' [the inset of
Fig.\ 19(a)].
Shown in Fig.\ 19(b) are the corresponding residues.
As for the branch 1', the dependence on $N$ is seen for $q/\pi\gsim 0.6$;
the branch 2' also has large dependence on $N$.
Consequently, we can interpret the above as follows: the branch 1'
is isolated except for $q\sim \pi$, and the branch 2' is the lower
edge of the continuum.
Thus the second gap vanishes for $q=\pi$, while an isolated branch
appears appreciably below the continuum for $q$ away from $q=\pi$.
Furthermore in the opening between them there is weak intensity
of the higher-order processes for $q/\pi\sim 0.5\pm0.2$.
The situation is quite different for larger values of $\delta$
($\gsim 0.15$), for which there appears an obvious second gap
even at $q=\pi$.
However, the second branch belongs to a continuum unlike the
cases of large $\alpha$.
\par

Summarizing, the appearance of a second gap or isolated branches
is dependent on the values of $\alpha$ and $\delta$.
The second gap, which is narrower than the first gap, may be
found at $q=\pi$ for CuGeO$_3$.
On the contrary, an isolated branch will be observed for $q\ne\pi$ for
$\alpha'$-NaV$_2$O$_5$.
\par

Finally, we mention a connection with the neutron experiments
under high pressure.\cite{pressure}
$T_{\rm SP}$ increases linearly with increasing pressure.\cite{Mori}
The lattice constant along the chain ($c$ axis) becomes slightly longer,
while the displacement of oxygen in the $ab$ plane is rather large.
The gap energy becomes twice larger (4.1meV) under 1.8GPa at 5K.
On the other hand, the dimerization for Cu atoms decreases under
high pressure.
\par

One possible interpretation of the above experiments, which seem mutually
contradictory at a glance, within the model of eq.\ (2.1) is that
$\alpha$ is enhanced almost up to the MG point by applying pressure.
If so, the gap becomes about twice larger even with reduced $\delta$
as in Fig.\ 16(b).
Since the abrupt enhancement of $\Delta$ is limited to the vicinity
of the MG point as in Fig.\ 11(b), this fact may be an evidence
of the strong frustration.
Anyway, more detailed experiments and reliable theoretical knowledge
on the relationship between the lattice structure and the model
parameters are needed to settle this issue.
\par

\section{Summary}
Using the exact diagonalization and the recursion methods,
we have investigated dynamical as well as static properties of the
1D $J$-$J'$ model with coupling alternation, and compared with
experiments of inorganic spin-Peierls compounds, CuGeO$_3$
and $\alpha'$-NaV$_2$O$_5$.
First, we have discussed the dynamical properties of the $J$-$J'$
model without coupling alternation.
The main points are:
\par
(1) For $\alpha<\alpha_{\rm c}$, the characteristics ([a]-[c] in
\S3.1 and \S3.2) of the excitation spectrum are basically the same
with those of the Heisenberg model.
\par
(2) For $\alpha\sim\alpha_{\rm c}$, the spectrum resembles that
of the Haldane-Shastry model.
\par
(3) As $\alpha$ further increases beyond $\alpha_{\rm c}$,
there appear qualitatively different features from the Heisenberg model,
summarized as [1]-[4] in \S3.3.
\par
Next, based on these results, we have studied a variety of features
of the above SP compounds, particularly in view of the inelastic
neutron scattering experiments.
We itemize noticeable points below:
\par
(4) Most characteristics of the complete neutron spectrum
at 10K are quantitatively reproduced by the 1D $J$-$J'$ model with
$\alpha=0.4$-0.45.
Some experimental aspects cannot be explained, unless $\alpha$ is
at least somewhat larger than $\alpha_{\rm c}$.
\par
(5) Using $\alpha=0.45$ (0.4), the strength of exchange coupling
is evaluated at $J\sim 180$K, consistent with the estimations by
other means.
The dimerization parameter of the exchange coupling $\delta$ is
found extremely small as $\sim 0.001$ (0.005).
\par
(6) We have investigated the isolated branch appearing below the
continuum or the double-gap structure for CuGeO$_3$ and
$\alpha'$-NaV$_2$O$_5$.
The features of the second gap will be different between these two
compounds.
\par
(7) Pressure effect on CuGeO$_3$ is possibly interpreted as
the increase of the frustration.
\par

Our estimation of $\alpha$ is somewhat larger than the ones by
$\chi(T)$, $\alpha\sim 0.35$.\cite{Kluemper}
Some causes of this discrepancy can be thought of.
For example, evaluations are made at different temperatures
between the two means.
Our estimation does not allow for the temperature effect.
There is quantitative ambiguity in neutron experiments at the
present stage.
Leaving aside such a quantitative point, we would like to emphasize
that the effect of the large frustration beyond $\alpha_{\rm c}$
is essential for various unique properties of CuGeO$_3$.
\par

There remain many significant issues left for further studies.
For example, although we treat a pure 1D model because the SP
transition itself proves good one dimensionality, a relatively
large magnetic dispersion perpendicular to the 1D chain was
observed in CuGeO$_3$.\cite{Nishi}
How do we interpret it?
What effect does the large frustration have on the elemental
substitutions?
Furthermore, phonon softening has never been discovered
in CuGeO$_3$,\cite{Kuroe} which has always been a driving force
of the organic SP transition so far.
Does the large frustration shoulder such a role, as conjectured
by B\"uchner \etal?\cite{Buechner}
To this end, the accumulation of experimental data on
$\alpha'$-NaV$_2$O$_5$ will be important.
\par\bigskip
\centerline{\bf Acknowledgments}
\par\smallskip
The authors are grateful to Yoshio Kuramoto and Yusuke Kato
for useful discussions, and to Masatoshi Arai and Masakazu
Nishi for providing them with the neutron data.
They thank the referee for pointing out their mistakes in
the original manuscript.
They learned useful techniques from the diagonalization
programs, TITPACK vers.2 by Hidetoshi Nishimori.
A part of the computations was done by using the facilities
of the Supercomputer Center, Institute for Solid State Physics,
University of Tokyo.
This study is partly supported by Grant-in-Aids for Scientific
Research on Priority Areas, ``Anomalous Metallic States
near the Mott Transition" and ``Physical Properties of Strongly
Correlated Electron Systems" and for Encouragement of Young Scientists,
given by the Ministry of Education, Science, Sports and Culture.
\par
\appendix
\section{Single-Mode Approximation}
In this Appendix, we summarize the properties of the single-mode
approximation (SMA)\cite{Feynman,HB} in the $J$-$J'$ model.
As a kind of variational trial state for a spin-excited state
with momentum $q$ we consider
\begin{equation}
|\psi_q\rangle=S_q^z|0\rangle,
\end{equation}
where $|0\rangle$ is the ground state---in this case exact.
The excitation energy with respect to $\psi_q$ is written as
\begin{eqnarray}
E_q-E_0&=&\frac{\langle \psi_q|{\cal H}|\psi_q\rangle}
{\langle \psi_q|\psi_q\rangle}-E_0
= \frac{\langle 0|\left[S_{-q}^z,\left[{\cal H},S_q^z\right]\right]/2
|0\rangle}{\langle 0|S_{-q}^z S_q^z|0\rangle} \nonumber \\
&=&\frac{K^{(1)}(q)}{S(q)}=\frac{\int d\omega \omega S(q,\omega)}
{\int d\omega S(q,\omega)}\equiv \bar\omega_q,
\end{eqnarray}
where $K^{(1)}(q)$ is the first frequency moment.
As seen in the last expression of eq.(A.2), $\bar\omega_q$ can be
regarded as a kind of average of $\omega$ [the distribution function
is $S(q,\omega)$], besides as a variational excitation energy.
Hence the SMA becomes good when the distribution of $S(q,\omega)$
concentrates on a small range of $\omega$;
in particular if the weight of $S(q,\omega)$ is restricted to a
certain sole value, the SMA becomes exact.
We will prove that a pure single mode is realized for the exactly
soluble cases at $q=\pi/2$ in the next Appendix.
Inversely, when the distribution of $S(q)$ spreads over
a wide range of $\omega$ evenly, the SMA is poor as the
approximation of the lowest mode.
In the meantime, since $\bar\omega_q$ is an average, it is
an upper bound for the lowest spin excited mode.
Therefore, one can use the SMA for a proof of the gaplessness.
\par

For the Hamiltonian eq.\ (1.1), the first frequency moment is
calculated as
\begin{equation}
K^{(1)}(q)=-2J\left[(1-\cos q)S_1+\alpha(1-\cos 2q)S_2\right],
\end{equation}
where $S_i$ is the two point correlation function with respect to
the ground state:
\begin{equation}
S_i=\frac{1}{N}\sum_j\langle S_j^zS_{j+i}^z\rangle.
\end{equation}
Thus, by using the averages of the exact diagonalization, the SMA
can be estimated.
In Fig.\ 20 we depict the results for the $J$-$J'$ model,
thus obtained.
\par

For $\alpha=0$ and small $q$, the SMA curve is slightly higher than
both of the two curves surrounding the TSC; this is due to the HOS
processes as mentioned in \S3.1.\cite{Mueller}
For $\alpha=0.2$ the result is close to that of the HS model,
for which the SMA is estimated through the known analytical formulae
of $K^{(1)}(q)$\cite{Mucciolo} and $S(q):$\cite{GV}
\begin{equation}
\frac{\tilde K^{(1)}(q)}{J}=\frac{q}{16}(2\pi-q)
+\frac{1}{8}(q-\pi)^2\ln\left(1-\frac{q}{\pi}\right),
\end{equation}
\begin{equation}
\tilde S(q)=-\frac{1}{4}\ln\left(1-\frac{q}{\pi}\right).
\end{equation}
This tendency has been discussed in \S3.2.
\par

For the MG case, $K^{(1)}(q)=J(1-\cos q)/4$ and $S(q)=(1-\cos q)/4$,
therefore $\bar\omega_q=J$ independent of the value of $q$.
As will be shown in the next Appendix, the spectrum becomes purely
single at $q=\pi/2$; the SMA becomes exact then.
Similarly, the SMA is relatively good for $\alpha=0.45$ and $q\sim\pi/2$,
though not exact.
On the other hand, for $q=0$ or $\pi$ the SMA is poor as the
lowest excitation.
In the viewpoint of an average, however, the SMA indicates that there
exists sizable contribution above $\omega=J$ to balance the
lower excitation modes, as seen in Fig.\ 7(a).
\par

To discuss the existence of the gap quantitatively in the SMA,
larger system sizes are desired.
\par

\appendix
\section{Proof of a Single Mode for Exactly Soluble Cases}
In this Appendix, we describe a proof \`a la Majumdar and Ghosh\cite{MG}
that $S(q,\omega)$ at $q=\pi/2$ in the exactly soluble cases
($2\alpha+\delta=1$) is a pure single mode with a excitation energy
independent of system size.
The case of $\delta=0$ has been already discussed,\cite{Yu} and
a more compact proof was given of $\psi_{\rm e}$ being an
eigenstate.\cite{Caspers}
\par

If $S_q^z|\Psi_0\rangle$ becomes an eigenstate of ${\cal H}$,
the surviving contribution to the sum of eq.\ (2.4) is given only
by this eigenstate, due to the orthogonality of the eigenstates.
Therefore, what we have to do for the proof is to show
$S_q^z|\Psi_0\rangle$ is an eigenstate of ${\cal H}$, and then to
estimate its eigenvalue.
\par

The proof that $\psi_{\rm e}\equiv S_{q=\pi/2}^z|\Psi_0\rangle$ becomes an
eigenstate of ${\cal H}$ is obtained by an analogous treatment
with the proof of the ground state.\cite{MG}
Henceforth, we consider systems of $N=4I$ ($I$: integer).
Before going to the proof, we summarize notations and relations.
Like a singlet pair $[i,j]$ of eq.\ (2.3), we denote a triplet pair by
\begin{equation}
\{i,j\}=\alpha(i)\beta(j)+\beta(i)\alpha(j).
\end{equation}
Useful relations for this proof are as follows:\cite{MG}
\begin{eqnarray}
&&\frac{1}{2}\left(1-4{\mib S}_\ell\cdot{\mib S}_m\right)[\ell,m]
=2[\ell,m], \\
&&\frac{1}{2}\left(1-4{\mib S}_\ell\cdot{\mib S}_m\right)\{\ell,m\}
=0,
\end{eqnarray}
\vspace{-5mm}
\begin{eqnarray}
&&\quad\frac{1}{2}\left(1-4{\mib S}_\ell\cdot{\mib S}_m\right)[k,\ell][m,n]
=[\ell,m][n,k], \\
&&\frac{1}{2}\left(1-4{\mib S}_\ell\cdot{\mib S}_m\right)[k,\ell]\{m,n\}
\nonumber \\
&=&\frac{1}{2}\left(1-4{\mib S}_\ell\cdot{\mib S}_m\right)\{k,\ell\}[m,n]
=-[\ell,m]\{n,k\},\qquad
\end{eqnarray}
\begin{equation}
[k,\ell][m,n]+[k,n][\ell,m]+[k,m][n,\ell]=0,
\end{equation}
\begin{equation}
[k,\ell]\{m,n\}+[\ell,m]\{n,k\}+[m,n]\{k,\ell\}+[n,k]\{\ell,m\}=0.
\end{equation}
\par
First, we consider the case of $\delta=0$.
To make use of the above relations, we rewrite the Hamiltonian
eq.\ (1.1) as,
\begin{equation}
{\cal H}=\frac{1}{4}NJ(1+\alpha)-\frac{1}{2}J\tilde{\cal H},
\end{equation}
with $\tilde{\cal H}=\tilde{\cal H}_1+\tilde{\cal H}_2$ and,
\begin{eqnarray}
\tilde{\cal H}_1&=&\sum_{j=1}^N \frac{1}{2}(1-4{\mib S}_j\cdot
                                           {\mib S}_{j+1}), \\
\tilde{\cal H}_2&=&\alpha\sum_{j=1}^N \frac{1}{2}(1-4{\mib S}_j\cdot
                                           {\mib S}_{j+2}).
\end{eqnarray}
As a ground-state eigenfunction, we take a form of eq.\ (2.2);
substitution by $\psi_2$ makes no difference.
Then, the function of our concern is given as,
\begin{eqnarray}
&&\psi_{\rm e}= S_{q=\frac{\pi}{2}}^z|\psi_1\rangle
      =\frac{1}{\sqrt{N}}\sum_{j=1}^N \exp\left(i\frac{\pi}{2}r_j\right)
S_j^z|\psi_1\rangle
      \nonumber \\
    &=&\frac{1}{\sqrt{N}}\left(iS_1^z-S_2^z-iS_3^z+S_4^z+\cdots\right)
      [1,2]\cdots[N-1,N] \nonumber \\
    &=&\frac{1}{2\sqrt{N}}(1+i)\Bigl(\tau_1-\tau_3+\tau_5-\tau_7+\cdots
                         \nonumber \\
    && \makebox[10em]{}               +\tau_{N-3}-\tau_{N-1}\Bigr),
\end{eqnarray}
where
\begin{equation}
\tau_j=[1,2]\cdots[j-2,j-1]\{j,j+1\}[j+2,j+3]\cdots[N-1,N],
\end{equation}
with odd $j$.
\par

First of all, let us take up $\tilde{\cal H}\tau_1$, for example.
Applying $\tilde{\cal H}_1$ and $\tilde{\cal H}_2$ to
$\tau_1$ and $\tau_3$ and using eqs.\ (B.2)-(B.5), we obtain
\begin{eqnarray}
\tilde{\cal H}_1\tau_1 &&=(N-2)\tau_1
                 -[2,3]\{4,1\}[5,6]\cdots[N-1,N] \nonumber \\
              &&+\{1,2\}[4,5][6,3]\cdots[N-1,N] + \cdots  \nonumber \\
              &&+\{1,2\}[3,4]\cdots[N-2,N-1][N,N-3] \nonumber \\
              &&-[N,1]\{N-1,2\}[3,4]\cdots[N-3,N-2],\qquad\qquad
\end{eqnarray}
\vspace{-5mm}
\begin{eqnarray}
&&\tilde{\cal H}_1\tau_3 =(N-2)\tau_3
                -[2,3]\{4,1\}[5,6]\cdots[N-1,N] \nonumber \\
              &&-[1,2][4,5]\{6,3\}[7,8]\cdots[N-1,N]  \nonumber \\
              &&+[1,2]\{3,4\}[6,7][8,5][9,10]\cdots[N-1,N]+\cdots \nonumber \\
              &&+[1,2]\{3,4\}[5,6]\cdots[N-2,N-1][N,N-3],\qquad\qquad
\end{eqnarray}
\vspace{-5mm}
\begin{eqnarray}
\tilde{\cal H}_2\tau_1/\alpha&&=-[1,3]\{4,2\}[5,6]\cdots[N-1,N] \nonumber \\
            &&+[2,4]\{3,1\}[5,6]\cdots[N-1,N]       \nonumber \\
            &&-2\{1,2\}[3,5][6,4][7,8]\cdots[N-1,N]-\cdots \nonumber \\
            &&-2\{1,2\}[3,4]\cdots[N-3,N-1][N,N-2] \nonumber \\
            &&+[N-1,1]\{N,2\}[3,4]\cdots[N-3,N-2] \nonumber \\
            &&-[N,2]\{N-1,1\}[3,4]\cdots[N-3,N-2],\qquad\qquad
\end{eqnarray}
\vspace{-5mm}
\begin{eqnarray}
&&\tilde{\cal H}_2\tau_3/\alpha = [1,3]\{4,2\}[5,6]\cdots[N-1,N]  \nonumber \\
          &&-[2,4]\{3,1\}[5,6]\cdots[N-1,N]       \nonumber \\
          &&-[1,2][3,5]\{6,4\}[7,8]\cdots[N-1,N]  \nonumber \\
          &&+[1,2][4,6]\{5,3\}[7,8]\cdots[N-1,N]  \nonumber \\
          &&-2[1,2]\{3,4\}[5,7][8,6][9,10]\cdots[N-3,N-2]-\cdots\nonumber \\
          &&-2[N-1,1][2,N]\{3,4\}[5,6]\cdots[N-3,N-2].\qquad\qquad
\end{eqnarray}
To begin with, we consider terms appearing in
$\tilde{\cal H}_2(\tau_1-\tau_3)$.
The first two terms in eqs.(B.15) and (B.16),
\begin{displaymath}
-2\left([1,3]\{4,2\}+[4,2]\{1,3\}\right)[5,6]\cdots[N-1,N],
\end{displaymath}
are transformed into
\begin{equation}
2\left([3,4]\{2,1\}+[2,1]\{3,4\}\right)[5,6]\cdots[N-1,N],
\end{equation}
by using eq.\ (B.7).
We assign (or should say return) the first term of eq.\ (B.17) to
$\tilde{\cal H}_2\tau_1$ instead of the first two terms of eq.\ (B.15),
and the second term of eq.\ (B.17) to $\tilde{\cal H}_2\tau_3$
instead of the first two terms of eq.\ (B.16).
Similarly, we can substitute
\begin{displaymath}
2[N-1,N]\{1,2\}[3,4]\cdots[N-3,N-2],
\end{displaymath}
for the last two terms of eq.\ (B.15) by considering
$\tilde{\cal H}_2(-\tau_{N-1}+\tau_1)$.
Next, we apply the relation eq.\ (B.6) to the two singlet pairs on which
$\tilde{\cal H}_2$ has been operated (namely pairs of irregular order)
in residual terms of eq.\ (B.15).
This substitution generates diagonal terms ($\tau_1$) and the same
kind of terms appearing in $\tilde{\cal H}_1\tau_1$ [from the third
term to the last but one in eq.\ (B.13)].
The residual off-diagonal terms, namely the second and the last ones
in eq.\ (B.13) cancel out with the corresponding terms in eq.\ (B.14)
and $\tilde{\cal H}_1\tau_{N-1}$, respectively.
Consequently, eqs.\ (B.13) and (B.15) are modified as
\begin{eqnarray}
(\tilde{\cal H}_1&+&\tilde{\cal H}_2)\tau_1
= (N-2)\tau_1+\alpha N\tau_1 \nonumber \\
&+& (1-2\alpha)\Bigl(\{1,2\}[4,5][6,3][7,8]\cdots[N-1,N]+\cdots \nonumber \\
&&+ \{1,2\}[3,4]\cdots[N-2,N-1][N,N-3]\Bigr)\qquad
\end{eqnarray}
If we put $\alpha=1/2$, the off-diagonal part in eq.\ (B.18) vanishes,
resulting in
\begin{equation}
\tilde{\cal H}\tau_1=\left(\frac{3}{2}N-2\right)\tau_1.
\end{equation}
\par

As for $\tau_3,\tau_5,\cdots$, the same form with eq.\ (B.19) can
be derived in the same manner.
Then, by considering eqs.\ (B.11) and (B.8), the following eigenvalue
equation can be obtained for the Majumdar-Ghosh model,
\begin{equation}
{\cal H}\psi_{\rm e}=\Bigl(-\frac{3}{8}N+1\Bigr)J\psi_{\rm e}.
\end{equation}
Since the ground-state energy is $E_0=-3NJ/8$, the eigenvalue
of $\psi_{\rm e}=S_{\pi/2}^z\psi_1$ is above $E_0$ by $J$,
independent of the system size $N$.
\par
This proof can be extended to the case of the exactly-soluble
dimerized $J$-$J'$ model eq.\ (2.1) with $2\alpha+\delta=1$.
Note that $\psi_2$ cannot be chosen in eq.\ (B.11) here.
Then, the eigenvalue equation corresponding to eq.\ (B.20) becomes
\begin{equation}
{\cal H}\psi_{\rm e}
 =\left[-\frac{3}{4}(1-\alpha)N+2(1-\alpha)\right]J\psi_{\rm e}.
\end{equation}
Since the first term is the ground-state energy in this case,\cite{SS}
the energy increment is $2(1-\alpha)J$.

\vfil\eject
\par
\vfil\eject
\begin{figure}
\caption{
$S(q,\omega)$ of the Heisenberg model for $N=26$.
The poles are situated at the center of the spheres.
The intensity of each pole is proportional to the {\it volume}
of the sphere to emphasize the poles with weak intensity.
The boundaries of the two-spinon continuum, $\omega_\ell(q)$ and
$\omega_{\rm u}(q)$, are shown by solid lines.
The lower edge of the higher-order-spinon continuum is indicated
by a dashed line ($N=26$), which is strongly dependent on $N$.
The lowest quintet level for each $q$ ($N=14$) is marked with
solid diamond.
}
\label{fig:1}
\end{figure}

\begin{figure}
\caption{
(a) Dependence on system size of residues which belong to the
lowest two two-spinon branches for the Heisenberg model.
The lowest (second-lowest) branch is indicated by solid symbols and 1'
(open symbols and 2').
(b) Dependence on $N$ of the poles belonging to the lowest four
two-spinon branches (1'-4') for $\alpha=0$.
Symbols have the same meanings with (a).
The inset shows the dependence on $N$ of poles at $q=\pi$ as a
function of $1/N^2$.
}
\label{fig:2}
\end{figure}

\begin{figure}
\caption{
(a) Total weight of residues outside the main continuum versus $q/\pi$
for some values of $\alpha(\le\alpha_{\rm c})$ and $N$.
The poles by higher-order processes are located outside of the
TSC for the present values of $q$ (see Fig.\ 1).
(b) Maximum values of the same quantity as a function of $\alpha$
for $0\le q/\pi\le 0.5$.
The arrows near the vertical axes indicate the direction of the dependence
on $N$ to the thermodynamic limit around their positions.
}
\label{fig:3}
\end{figure}
\begin{figure}
\caption{
$S(q,\omega)$ for $\alpha=0.2$ and $N=26$.
The intensity of each pole is proportional to the {\it volume}
of the sphere, and in the same scale with Fig.\ 1.
The boundaries of the two-spinon continuum for the HS model,
$\tilde\omega_\ell(q)$ and $\tilde\omega_{\rm u}(q)$, are shown by
solid lines.
}
\label{fig:4}
\end{figure}
\begin{figure}
\caption{
(a) Static susceptibility as a function of $q$ for
$\alpha=0, 0.2, 0.25, 0.4, 0.45, 0.5$ from bottom to top, respectively.
Lines (except solid line) indicate polynomial fit of 26-site
data as a guide for the eye.
The solid line is the value for the HS model given by eq.\ (3.6).
The arrow and cross on the vertical axis indicate the exact value both
for the Heisenberg model and the HS model: $1/2\pi^2$.
The inset is a further magnification of small-$q$ regime to recognize
singular behavior of the Heisenberg model.
(b) Magnified figure of (a) for small $q$.
Symbols are common in (a) and (b).
}
\label{fig:5}
\end{figure}
\begin{figure}
\caption{
(a) $K^{(2)}(q)$ and (b) $K^{(3)}(q)$ for some values of $\alpha$.
Symbols are common between (a) and (b).
Lines (except solid line) are guide for the eye.
The solid lines in both panels are of the HS model given by
eqs.\ (3.7) and (3.8), respectively.
}
\label{fig:6}
\end{figure}
\begin{figure}
\caption{
(a) Residues for the MG model.
The system with $N=24$ is shown to include the wave number $\pi/2$.
(b) $S(q,\omega)$ for the MG model ($N=26$).
In both (a) and (b), both contributions from $\Psi_+$ and $\Psi_-$
are plotted together, and the weight is indicated by the {\it area}
of the circles.
}
\label{fig:7}
\end{figure}
\begin{figure}
\caption{
(a) Dependence on $N$ of residues for the first (1') and second (2')
lowest dominant branches as a function of $q$.
(b) The same for the third (3') and fourth (4') dominant branches.
Both contributions from $\Psi_+$ and $\Psi_-$ are plotted together.
(c) Dependence on $N$ of the pole positions corresponding to the
residues in (a) and (b).
The cross and arrow on the right vertical axis show the extrapolated
pole position of the branch 1' in the thermodynamic limit, which means
the gap of the MG model: $\Delta/J \sim0.233$ in our estimation by the
fourth-order polynomial fit.
The inset shows the movement of the poles at $q=\pi$ as the system
size changes.
Since the spectrum is symmetric with respect to $q=\pi/2$,
we only show $0.5\le q/\pi\le 1.0$.
}
\label{fig:8}
\end{figure}
\begin{figure}
\caption{
Residues at $q=\pi$ as a function of $\omega/J$ for some system sizes.
The connected points with numbers (1' etc.) correspond to
the poles in Fig.\ 8.
The arrow indicates the peak near the upper edge of the main continuum.
Since the number of coefficients used in the recursion method
is different for different systems, namely 30 for $N\ge 22$ and 100
for $N\le 20$, many poles with weak intensity appear for the small
systems.
}
\label{fig:9}
\end{figure}
\begin{figure}
\caption{
$S(q,\omega)$ of $N=26$ for (a) $\alpha=0.35$, (b) 0.4 and
(c) 0.45.
In this size the ground state is $Q=\pi$.
The intensity is proportional to the {\it area} of the circle,
and in the same scale with Fig.\ 7(b).
}
\label{fig:10}
\end{figure}
\begin{figure}
\caption{
(a) The ground-state energy, (b) the lowest excited energy at
$q=\pi/2$ and the gap, each as a function of $\alpha$.
Each value is extrapolated from the finite-size data ($N=8$-26)
by using some orders of polynomial fit.
The values for the HS model are indicated by crosses and arrows
on the vertical axes.
The energy and the gap were previously estimated
in some references.\cite{TH}
}
\label{fig:11}
\end{figure}
\begin{figure}
\caption{
$S(q)$ for some large values of $\alpha$.
Each line connects the data of $N=26$ except for $\alpha=0.5$.
The arrows along the right vertical axis indicate the directions
of the dependence on $N$ of $S(\pi)$ for $N\rightarrow\infty$.
}
\label{fig:12}
\end{figure}
\begin{figure}
\caption{
Comparison of $S(q)$ between the diagonalization results
for $\delta=0$ and the experimental results for 10K and 20K,
which are taken from Fig.\ 2 of ref.\ 19.
The arrows along the right vertical axis indicate the directions
of the dependence on $N$ for $N\rightarrow\infty$.
}
\label{fig:13}
\end{figure}
\begin{figure}
\caption{
$S(q,\omega)$ at a fixed value of $\delta=0.05$ for three values
of $\alpha$, (a) 0, (b) 0.2 and (c) 0.4.
$N=26$.
In each figure, the intensity is proportional to the {\it area}
of the circle, and in the same scale with Figs.\ 7(b) and 10.
}
\label{fig:14}
\end{figure}
\begin{figure}
\caption{
Lowest excitation at $q=\pi/2$ as a function of $\delta$
for some values of $\alpha$.
Shown data are extrapolated from finite-size results of $N=8$-24
with using eq.\ (2.9) and $m=1$.
The values for $\delta=0$ correspond to Fig.\ 11(b).
The arrow on the vertical axis indicates the value of
des Cloizeaux-Pearson, $\pi/2$.
Shaded lines are guide for the eye.
The dash-dotted line shows the exactly soluble case, as given
in Appendix B.
}
\label{fig:15}
\end{figure}

\begin{figure}
\caption{
(a) Gap of the model eq.\ (2.1) as a function of $\delta$ for some
values of $\alpha$.
Each value is extrapolated from the results of $N=8$-26 by using
eq.\ (2.9).
The arrow on the vertical axis indicates the gap for the MG model.
If we fit for each $\alpha$ with power functions, the resultant
powers are 0.71 ($\alpha=0.2$), 0.61 (0.35), 0.58 (0.4) and
0.51 (0.45).
Dashed (dash-dotted) line indicates a plausible value of
$\alpha$'-NaV$_2$O$_5$ for $\Delta=85$K and $J=441$K\cite{Weiden}
($\Delta=98$K and $J=560$K\cite{Ueda}).
The data for small $\delta$ and/or $\alpha$ are omitted, because
reliable results cannot be obtained by extrapolations.
Size of the symbols and width of the lines roughly represent the
inaccuracy of the extrapolation.
(b) Magnification of (a) for a small-$\delta$ regime.
Dash-dotted (dashed) line indicates a value of CuGeO$_3$
($\Delta=24$K) for $J=180$K (160K). See text.
}
\label{fig:16}
\end{figure}

\begin{figure}
\caption{
Total energy of the model eq.\ (2.1) as a function of $\delta$
for some values of $\alpha$.
Each value is extrapolated from the results of $N=8$-26 by using
eq.\ (2.9).
If we fit the data for each $\alpha$ with power functions in this
range of $\delta$, the resultant powers are 1.42 ($\alpha=0$),
1.30 (0.2), 1.13 (0.35), 1.07 (0.4) and 1.03 (0.45).
Dash-dotted line indicates the soluble case of $2\alpha+\delta=1$.
}
\label{fig:17}
\end{figure}
\begin{figure}
\caption{
(a) Dependence on system size of the pole positions of the lowest
two branches.
The parameters roughly correspond to CuGeO$_3$.
The subtle asymmetry with respect to $q=\pi/2$ is due to numerical
round-off errors.
The inset shows the dependence on $N$ at $q=\pi$ for $N=8$-26,
fitted by eq.\ (2.9).
(b) Dependence on $N$ of the corresponding residues.
The arrow on the right axis indicates the extrapolated value
for $N\rightarrow\infty$ of the branch 1'.
}
\label{fig:18}
\end{figure}

\vfil\eject
\begin{figure}
\caption{
(a) Dependence on $N$ of the pole positions for the lowest
three {\it dominant} branches.
The parameters roughly correspond to $\alpha'$-NaV$_2$O$_5$.
Concerning the branch 3', we show only for $q/\pi\gsim 0.4$.
By `+' we represent poles with tiny weight; to avoid mess we plot
only for $q/\pi\gsim 0.6$ and under the branch 3'.
The inset shows the dependence on $N$ at $q=\pi$ for $N=8$-26
with polynomial fit.
(b) Dependence on $N$ of the corresponding residues.
}
\label{fig:19}
\end{figure}
\begin{figure}
\caption{
Lowest spin excited modes for the 1D $J$-$J'$ model obtained
by the SMA.
For the MG case (dash-dotted line), $\omega=J$.
Otherwise, the diagonalization results are used.
Dotted lines are guide for the eye.
For comparison for $\alpha=0$, the exact lower (dCP) and upper
edges of the TSC are shown with dashed lines.
The lower edge for $\alpha=0.45$ obtained by the recursion method
($N=20$-26) is indicated by cross and plus.
The SMA and the lowest edge for the HS model is indicated by
solid and long dash-dotted lines, respectively.
The extrapolated values of the gap (exact) are shown by arrows
on the vertical axis for three values of $\alpha$.
}
\label{fig:20}
\end{figure}

\par
\vfil\eject

\begin{thebibliography}{99}
\bibitem{notetrans} Defining a translation operator $T$
as $TS^\mu_jT^\dagger=S^\mu_{j+1}$ ($\mu=x,y,z$), we classify
the eigenstates of ${\cal H}$ by the index $Q$, where $e^{iQ}$ is
the eigenvalue of $T$.
In the present case, the ground state of a 4$N$- (4$N$+2-) system
belongs to $Q=0$ ($\pi$).
\bibitem{notebz} In the cases of $\delta\ne 0$ a reduced Brillouin
zone of [$-\pi/2, \pi/2$] should be used.
However, since the intensity is not symmetric with respect to $\pi/2$,
we use the original zone for comparison with undimerized cases.
\bibitem{notebigd} As for [3], the divergence at $q=\pi$ of $S(q)$
tends to be suppressed by $\delta$.
However, $\delta$ never enhances the intensity partially for the
lower edge of $q\sim\pi/2$.
\bibitem{NMR} For example, in Cu-NQR experiment,\cite{Yasuoka} the
resolution is roughly estimated at 0.04\AA, and the displacement
was not observed for CuGeO$_3$, for which the observed value by
neutron diffraction is $\sim 0.004$\AA.\cite{Hirota1}
\bibitem{notelog} As for the energy, logarithmic correction is
limited to the region of fairly small values of $\delta$, so that
the prediction by Cross and Fisher, $E\propto\delta^{4/3}$, is
observed even for $\alpha=0$.\cite{BonnerB,Soos}
These facts tell us that one has to take seriously the effect of
logarithmic terms into account for each quantity.
\end{thebibliography}
\end{document}